\RequirePackage{ifpdf}
\ifpdf 
\documentclass[pdftex]{sigma}
\else
\documentclass{sigma}
\fi

\begin{document}
\allowdisplaybreaks

\renewcommand{\PaperNumber}{032}

\FirstPageHeading

\ShortArticleName{Localized Induction Equation for Stretched
Vortex Filament}

\ArticleName{Localized Induction Equation\\ for Stretched Vortex
Filament}

\Author{Kimiaki KONNO~$^\dag$ and Hiroshi KAKUHATA~$^\ddag$}
\AuthorNameForHeading{K. Konno and H. Kakuhata}

\Address{$^\dag$~Department of Physics, College of Science and Technology, Nihon University,\\
$\phantom{^\dag}$~Tokyo 101-8308, Japan}
\EmailD{\href{mailto:konno@phys.cst.nihon-u.ac.jp}{konno@phys.cst.nihon-u.ac.jp}}

\Address{$^\ddag$~Toyama University, Toyama 930-8555, Japan}
\EmailD{\href{mailto:kakuhata@iis.toyama-u.ac.jp}{kakuhata@iis.toyama-u.ac.jp}}

\ArticleDates{Received October 05, 2005, in f\/inal form February
16, 2006; Published online March 02, 2006}

\Abstract{We study numerically the motion of the stretched vortex
f\/ilaments by using the localized induction equation with the
stretch and that without the stretch.}

\Keywords{localized induction equation; stretch; vortex
f\/ilament}

\Classification{35Q51; 37K15; 76B29}

\section{Introduction}

Konno and Kakuhata are interested in motion of the stretched
vortex f\/ilaments where the f\/ilaments are described by the
localized induction equation (LIE)~\cite{Konno_Kakuhata2005}. LIE
is given by
\begin{gather}
    {\boldsymbol X}_t= {\boldsymbol X}_s \times  {\boldsymbol X}_{ss},
\label{eq:LIE}
\end{gather}
in which ${\boldsymbol X}$ is the position vector $(X,Y,Z)$, $s$
is the arclength and $t$ is the time. This LIE was f\/irst
obtained by Arms and Hama for the thin vortex f\/ilament with the
localized induction approximation~\cite{Hama_Ames}. Their original
equation is derived as
\begin{gather}
{\boldsymbol X}_t={{\boldsymbol X}_s \times {\boldsymbol X}_{ss}
\over |{\boldsymbol X}_s|^3}. \label{eq:LIE_s}
\end{gather}
Under the condition $|{\boldsymbol X}_s|=1$, (\ref{eq:LIE_s}) is
reduced to (\ref{eq:LIE}). $|{\boldsymbol X}_s|$ is equal to the
local stretch of the f\/ilament  $l_s$ def\/ined
by~\cite{Konno_Kakuhata2003_2}
\begin{gather*}
l_s={{\rm d} l \over {\rm d} s} = |{\boldsymbol X}_s|,
\end{gather*}
where $l$ is the length along the f\/ilament. If $l_s =1$, there
is no stretch, if $l_s >1$, stretch and if $l_s <1$, then shrink.

If there is no stretch, one can def\/ine the tangent vector
 ${\boldsymbol t}=\partial {\boldsymbol X}/\partial s$
 with its magnitude~1 and use the Frenet--Serret scheme.
 Then one introduces the curvature and the torsion, and shows that equation~(\ref{eq:LIE})
 is equivalent to the nonlinear Schr\"odinger equation through the Hashimoto transformation~\cite{Hashimoto}.
For $l_s \ne 1$ $s$ does not mean the arclength but the parameter
along the f\/ilament. LIE is an integrable equation. Moreover, we
have shown that (\ref{eq:LIE_s})
 is integrable and has a hierarchy which is constructed by the recursion operators~\cite{Konno_Kakuhata2005_2}.

In this paper, in order to compare (\ref{eq:LIE_s}) with
(\ref{eq:LIE}) we will try to solve an initial value problem with
the same initial vortex f\/ilaments. The numerical scheme is given
in~\cite{Konno_Mitsuhashi_Ichikawa}.

In the next section we show the numerical method for
(\ref{eq:LIE_s}). In Section~3 numerical results are given. The
last section presents conclusion.

\begin{figure}[htb]
\centerline{\includegraphics[width=4.5cm]{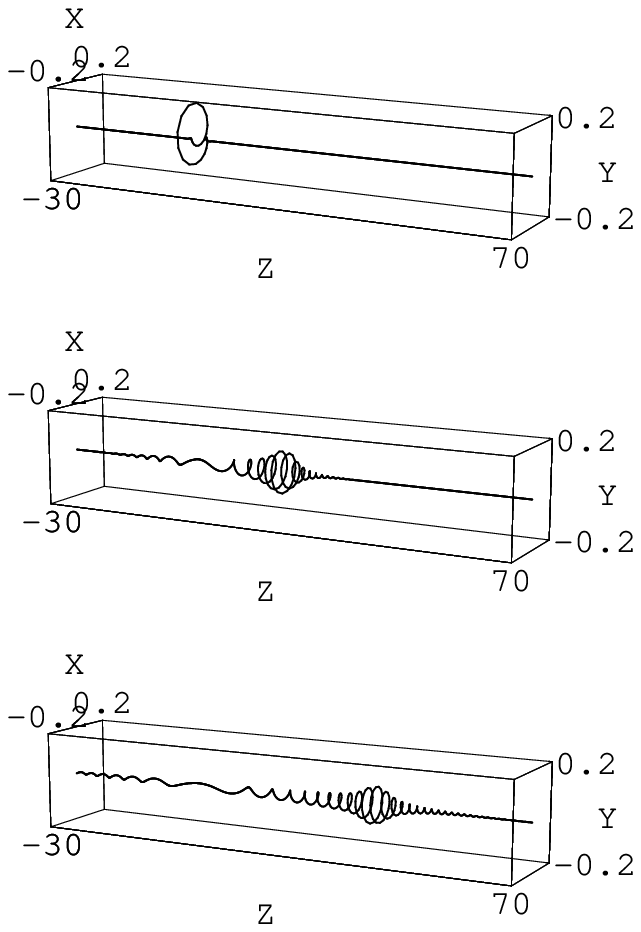}
\qquad\qquad
\includegraphics[width=3cm]{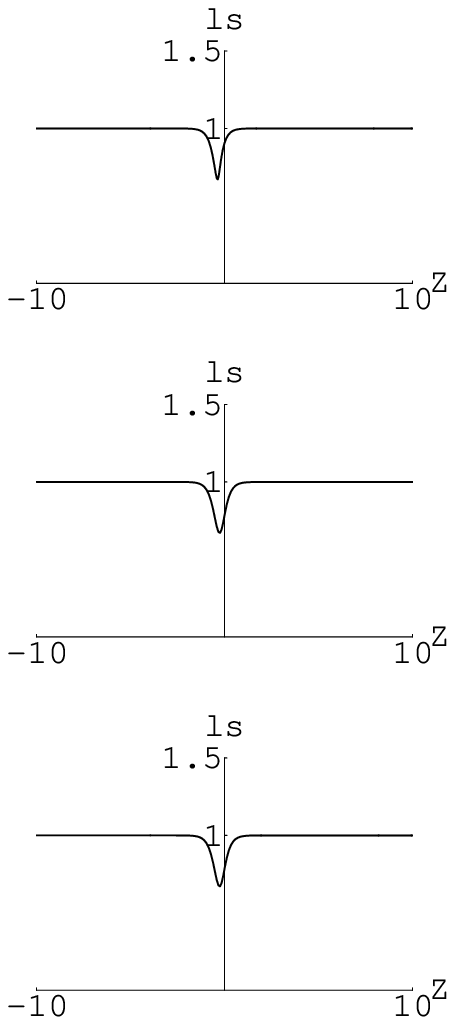}}
\vspace{-1mm}

\caption{Time evolution (left) and local stretch (right)  for the
stretched vortex soliton (\ref{eq:int_vorex}) of (\ref{eq:LIE}) at
$ t = 0,\, 3,\,6$ for $A=0.6$ and $\lambda =1.5+{\rm
i}$.\label{fig:vortex}} \vspace{-1mm}
\end{figure}

\section{Numerical method}

In order to solve (\ref{eq:LIE_s}), we use the following implicit
f\/inite dif\/ference equation~\cite{Konno_Mitsuhashi_Ichikawa},
as an example, for the $x$ component as
\begin{gather}
{X_{j+1,i}-X_{j,i} \over \Delta t} ={1 \over
L_s^3}\left[{Y_{j,i+1}-Y_{j,i}\over \Delta
s}{Z_{j+1,I+1}-2Z_{j+1,i}+Z_{j+1,i-1}\over \Delta s^2}\right.
\nonumber\\
 \left.\phantom{{X_{j+1,i}-X_{j,i} \over \Delta t}=}{}
-{Z_{j,i+1}-Z_{j,i}\over \Delta
s}{Y_{j+1,i+1}-2Y_{j+1,i}+Y_{j+1,i-1}\over \Delta s^2}\right],
\label{eq:numericalmethod}
\end{gather}
where
\begin{gather*}
L_s=\sqrt{{(X_{j,i+1}-X_{j,i})^2+(Y_{j,i+1}-Y_{j,i})^2+(Z_{j,i+1}-Z_{j,i})^2
\over \Delta s^2}}.
\end{gather*}
The subscripts $j$ and $i$ refer to the time $t=j \Delta t$ and
$s=i\Delta s$, respectively. (\ref{eq:numericalmethod}) and $Y$
and $Z$ components can be written as the following matrix form
\begin{gather}
D_{j,i}H_{j+1,i+1}+(1-2D_{j,i})H_{j+1,i}+D_{j,i}H_{j+1,i-1}=H_{j,i},
\label{eq:threeFold}
\end{gather}
where
\begin{gather*}
 H_{j,i}=\left( \begin{array}{l}
     X_{j,i} \\
     Y_{j,i} \\
     Z_{j,i} \\
    \end{array}
    \right),
\qquad
 D_{j,i}=\left( \begin{array}{lll}
     0 & c_{j,i} & -b_{j,i}\\
     -c_{j,i} & 0 & a_{j,i}\\
     b_{j,i} & -a_{j,i} & 0\\
    \end{array}
    \right)
\end{gather*}
and
\begin{gather*}
 a_{j,i}={\Delta t \over L_s^3 \Delta s^3}(X_{j,i+1}-X_{j,i}),\qquad
b_{j,i}={\Delta t \over L_s^3 \Delta s^3}(Y_{j,i+1}-Y_{j,i}),\\
c_{j,i}={\Delta t \over L_s^3 \Delta s^3}(Z_{j,i+1}-Z_{j,i}).
\end{gather*}
To solve (\ref{eq:threeFold}), we introduce a recurrence formula
\begin{gather}
H_{j+1,i}=E_{j,i}H_{j+1,i+1}+F_{j,i} \label{eq:recurrsion}
\end{gather}
where
\begin{gather}
 E_{j,i} =-(1-2D_{j,i}+D_{j,i}E_{j,i-1})^{-1}D_{j,i},\nonumber\\
 F_{j,i} = (1-2D_{j,i}+D_{j,i}E_{j,i-1})^{-1}(H_{j,i}-D_{j,i}F_{j,i-1}),
\label{eq:EFji}
\end{gather}
for $i =2, 3, \ldots, N$ and
\begin{gather}
E_{j,1} =-(1-2D_{j,1})^{-1}D_{j,1},\qquad F_{j,1} =
(1-2D_{j,i})^{-1}(H_{j,1}-D_{j,1}H_{j,0}). \label{eq:EFj1}
\end{gather}
Here we take the boundary condition
\begin{gather}
\displaystyle H_{j,0}=\left( \begin{array}{c}
     0\\
     0\\
     Z_{j,1}-\Delta s\\
    \end{array}
    \right).
\label{eq:EFj0}
\end{gather}
The values of all coef\/f\/icients (\ref{eq:EFji}) are recursively
determined by using (\ref{eq:EFji}) and (\ref{eq:EFj1}) with
(\ref{eq:EFj0}). Taking $X_{j,N+1}=0$, $Y_{j,N+1}=0$ and
$Z_{j,N+1}=Z_{j,N}+\Delta s$, we can solve the recurrence
formula~(\ref{eq:recurrsion}).

We have taken values of $\Delta s=1/8$ and $\Delta t=\Delta
s^3/32$. Number of steps $j$ is $12/\Delta t$. Numerical accuracy
is checked against the following aspects, i.e.\ i)~the steady
state vortex solution of~(\ref{eq:int_vorex}) with $A=1$
propagates without showing signif\/icant distortion and ii)~the
conserved quantity~\cite{Konno_Kakuhata2005_2} $\int (l_s -1){\rm
d} s$ conserves within the accuracy 5\,$\%$ for (\ref{eq:LIE}) and
10\,$\%$ for~(\ref{eq:LIE_s}).

\begin{figure}[t]
\centerline{\includegraphics[width=4.5cm]{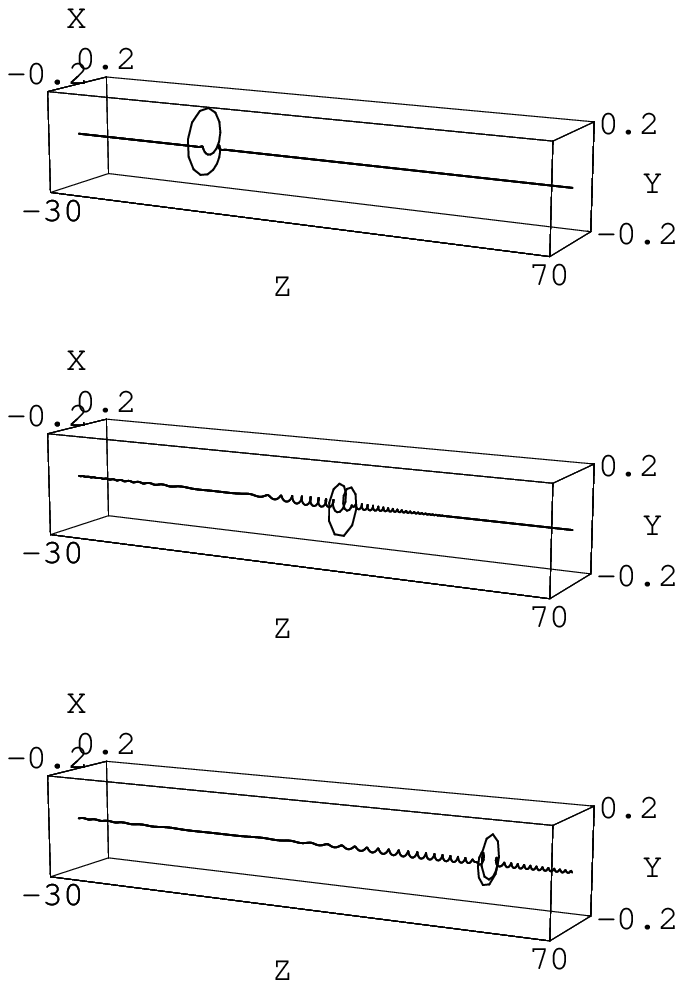}
\qquad\qquad
\includegraphics[width=3cm]{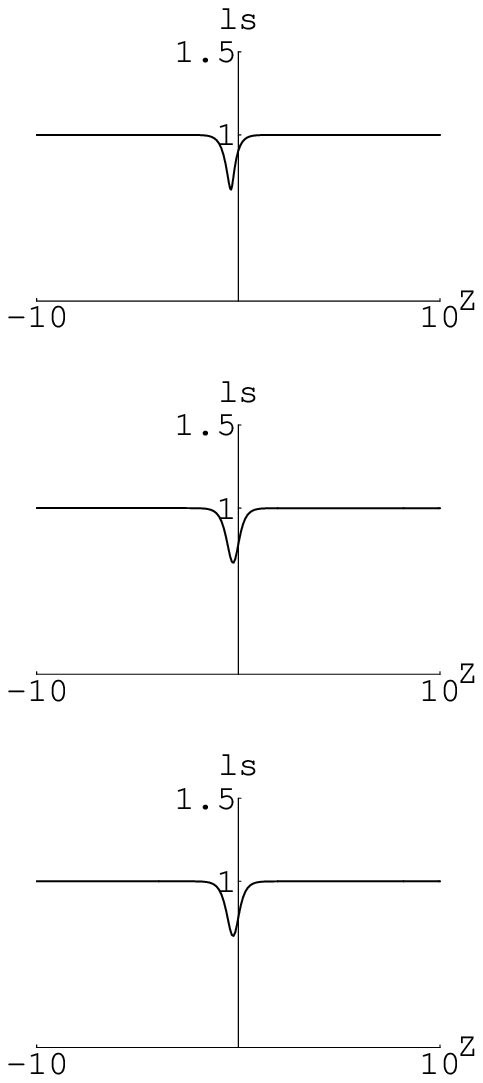}}
\vspace{-1mm}

\caption{Time evolution (left) and the local stretch (right) for
the stretched vortex soliton (\ref{eq:int_vorex}) of
(\ref{eq:LIE_s}) at $ t = 0,\,3,\,6$ for $A=0.6$ and
$\lambda=1.5+{\rm i}$.\label{fig:loopA06}} \vspace{-1mm}
\end{figure}

\begin{figure}[tb]
\centerline{\includegraphics[width=4.5cm]{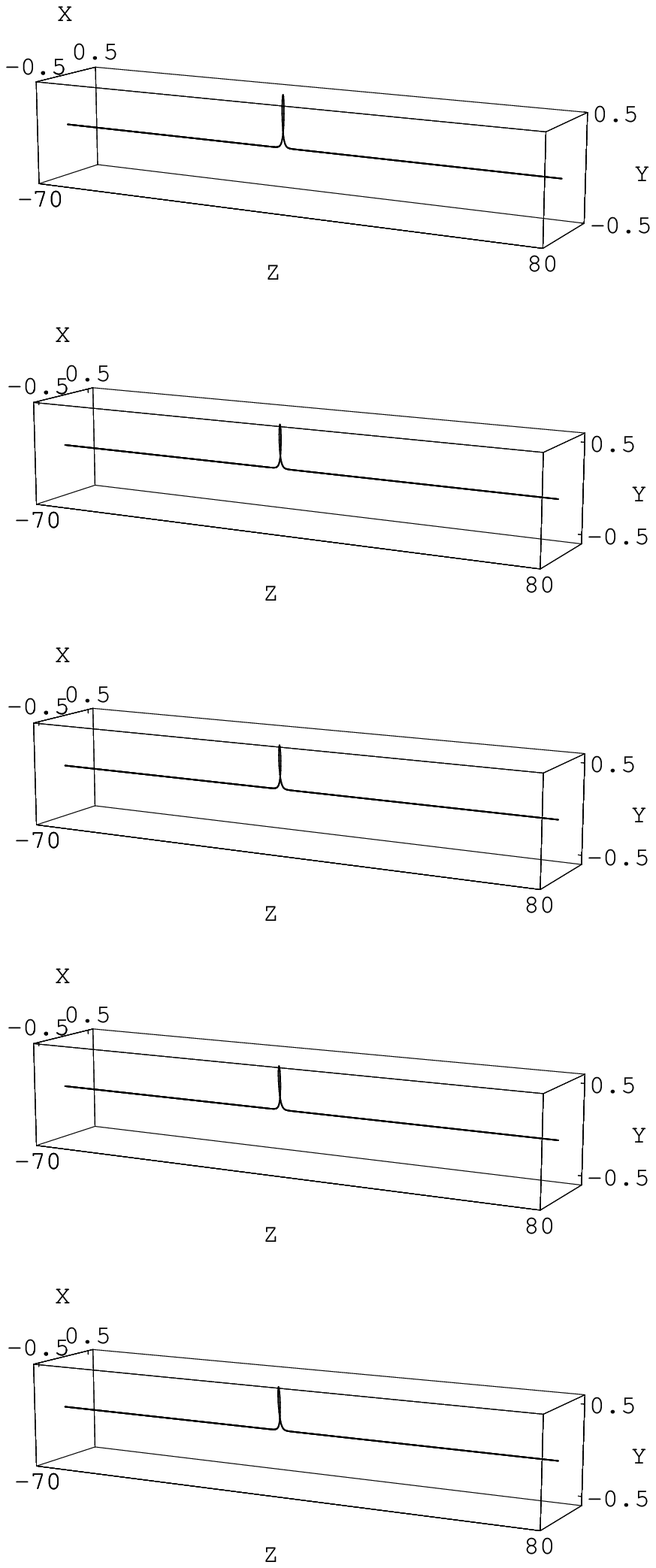}
\qquad\qquad
\includegraphics[width=3.0cm]{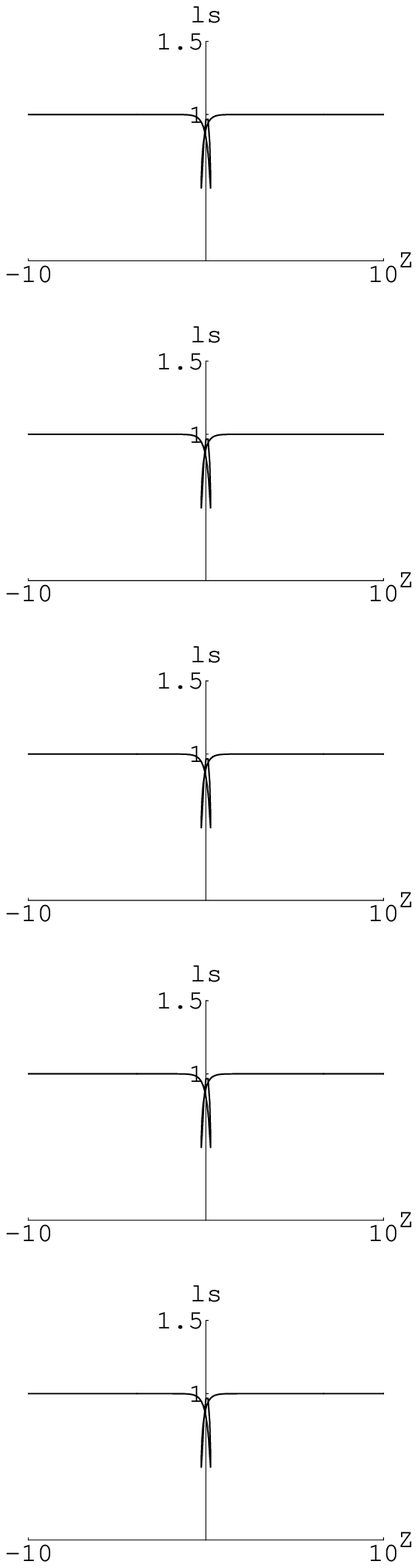}}
\vspace{-1mm}

\caption{Time evolution (left) and local stretch (right) for the
loop type of vortex soliton (\ref{eq:loop_vorex}) of
(\ref{eq:LIE}) at $ t = 0,\, 3,\, 6,\, 9,\, 12$ for
$A=0.5$.\label{fig:vortex_us}} \vspace{-1mm}
\end{figure}

\section{Initial value problem}

\subsection[Stretched vortex filament]{Stretched vortex f\/ilament}

As an initial prof\/ile of the stretched vortex f\/ilament we take
modif\/ied form of the one vortex soliton solution of
(\ref{eq:LIE}) as
\begin{gather}
 X=A{\lambda_I \over \lambda_R^2+\lambda_I^2}\sin (2 \Omega +\delta){\rm sech}(2 \Theta +\epsilon),\qquad
Y=-A{\lambda_I \over \lambda_R^2+\lambda_I^2}\cos (2 \Omega +\delta){\rm sech}(2 \Theta +\epsilon),\nonumber\\
 Z=s-{\lambda_I \over \lambda_R^2+\lambda_I^2}{\rm tanh}(2 \Theta+\epsilon),
\label{eq:int_vorex}
\end{gather}
where $A$ is the factor to represent the stretch and we take
$t=0$. If $A=1$, (\ref{eq:int_vorex}) is just the exact solution
without the stretch. $\Omega$,
 $\Theta$ and constants $\delta$, $\epsilon$ are given by
\begin{gather*}
\Omega=\lambda_R s- \omega_R t,\qquad \Theta=\lambda_I s-\omega_I
t,\qquad
 \tan \delta=-{2 \lambda_R \lambda_I \over \lambda_R^2-\lambda_I^2},
\qquad \epsilon=-{1 \over 2}\log {|c_0|^2 \over 4 \lambda_I^2},
\end{gather*}
where $c_0$ is a constant. The wave number $\lambda=\lambda_R+{\rm
i} \lambda_I$ and the frequency $\omega=\omega_R +{\rm i}
\omega_I$ are connected by the dispersion relation $\omega= 2
\lambda^2$. The local stretch is given by
\begin{gather*}
l_s^2=1+{4 \lambda_I^2 \over
\lambda_R^2+\lambda_I^2}\big(A^2-1\big)
      \left[{\rm sech}^2(2 \Theta+\epsilon)-{\lambda_I^2 \over \lambda_R^2+\lambda_I^2}
      {\rm sech}^4(2 \Theta+\epsilon)\right].
\end{gather*}
If $A=1$, there is no stretch. If $A \ne 1$, (\ref{eq:int_vorex})
shows the local stretch for $A>1$ and the local shrink for $A<1$.

Using LIE, we show the prof\/iles and the local stretch of the
vortex f\/ilament in Fig.~\ref{fig:vortex} for $A=0.6$ with
$\lambda=1.5+{\rm i}$ and $c_0=1$. We observe that, since the
f\/ilament has shrunk initially, then the vortex
 soliton deforms its prof\/ile and moves away with the strong radiation,
 however, the shrunk region remains in the same place as freezing the shrink~\cite{Konno_Kakuhata2005}.

For the case of the equation of motion (\ref{eq:LIE_s}), we
illustrate the prof\/iles and the local stretch of the f\/ilament
in Fig.~\ref{fig:loopA06} with the same parameters such as
$A=0.6$, $\lambda=1.5+{\rm i}$ and $c_0=1$. Comparing with LIE, we
observe that the vortex soliton takes a small modif\/ication and
moves fast with the weak radiation. The shrunk region remains in
the same place, which is similar to the case of LIE.

\begin{figure}[t]
\centerline{\includegraphics[width=4.5cm]{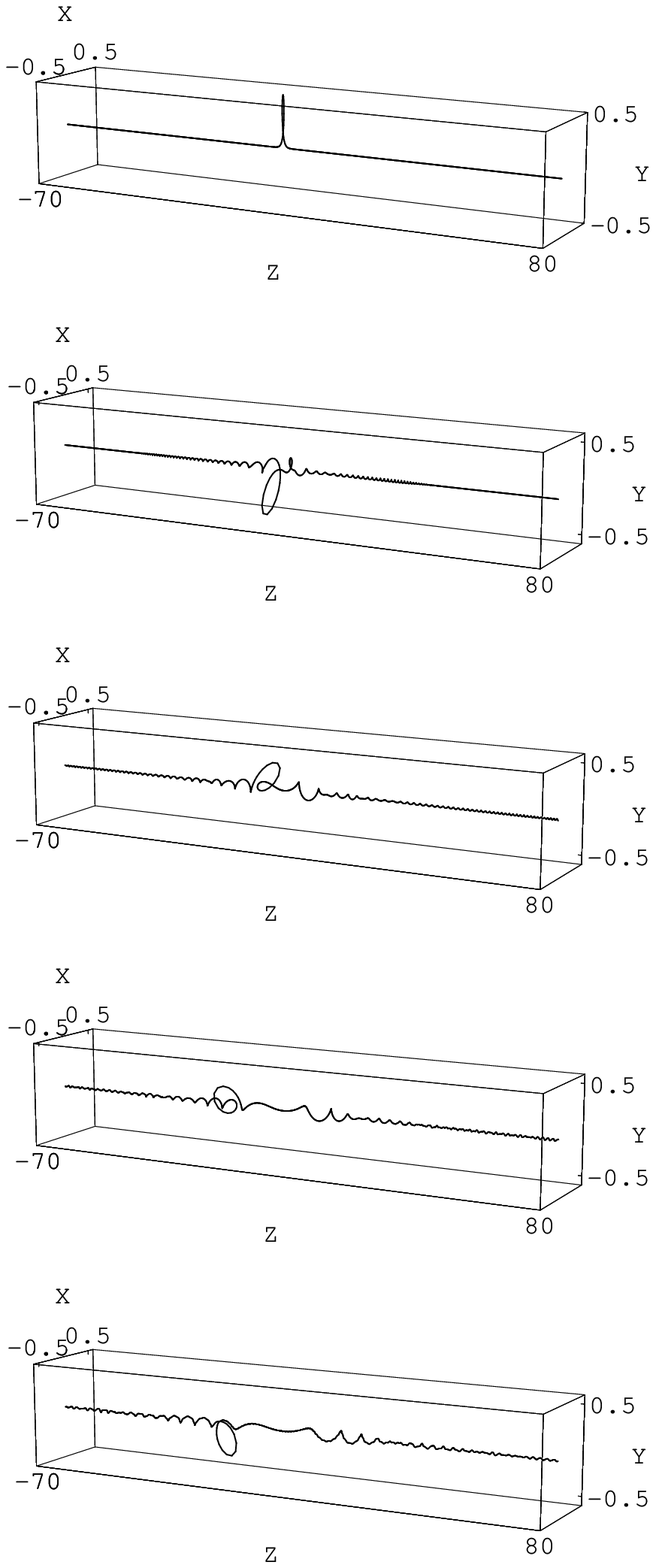}
\qquad\qquad
\includegraphics[width=3.0cm]{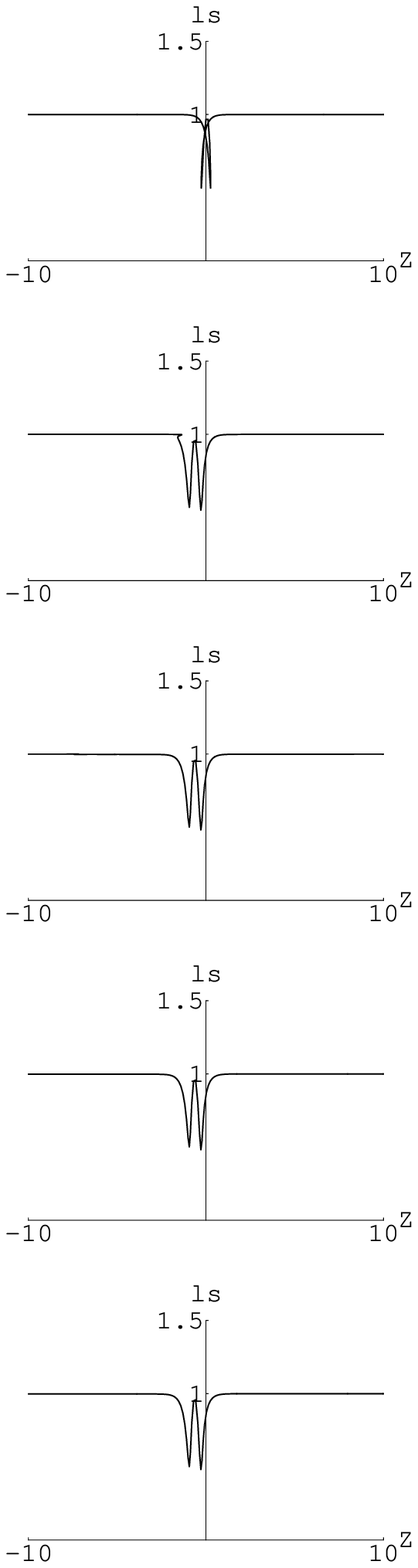}}
\vspace{-1mm}

\caption{Time evolution (left) and local stretch (right) for the
loop type of vortex soliton (\ref{eq:loop_vorex}) of
(\ref{eq:LIE_s}) at $ t = 0,\, 3,\, 6,\, 9,\, 12$ for
$A=0.5$.\label{fig:vortex_s}} \vspace{-1mm}
\end{figure}

\subsection[Stretched loop type vortex filament]{Stretched loop type vortex f\/ilament}

With the exact solution of (\ref{eq:LIE}) as
\begin{gather}
 X=-A\sin 4t\,{\rm sech} 2s,\qquad
  Y=A\cos 4t\,{\rm sech} 2s,\qquad
   Z=s-{\rm tanh} 2s,
\label{eq:loop_vorex}
\end{gather}
we consider the initial value problem by taking $t=0$. Here $A$ is
the factor to represent the stretch for the loop type of the
vortex f\/ilament. The local stretch is given by
\begin{gather*}
l_s^2=1+4\big(A^2-1\big){\rm sech}^2 2s\, {\rm tanh}^2 2s.
\end{gather*}

In Fig.~\ref{fig:vortex_us} we show the prof\/iles and the local
stretch of the loop type of the vortex f\/ilament for $A=0.5$ at
the instants $t=0,\,3,\,6,\,9,\,12$ by using (\ref{eq:LIE}). Since
the initial prof\/ile is the exact solution we observe that the
vortex f\/ilament rotates in the same place with a constant
angular velocity and that the shrunk region does not
move~\cite{Konno_Kakuhata2005}.

Using the evolution equation (\ref{eq:LIE_s}), we simulate the
vortex f\/ilament with the same parameter as $A=0.5$ and show the
prof\/iles and the local stretch at the instants
$t=0,\,3,\,6,\,9,\,12$ in Fig.~\ref{fig:vortex_s}. From the
prof\/iles, we can observe that new propagating vortex soliton is
produced by accompanied with the radiation. However, from the
local stretch, we f\/ind that shrunk region with double dips
remains in the nearly same place, which means that the mode with a
constant~$X$, a constant~$Y$ and two kinks for $Z-s$ excited in
the region.

\section{Conclusion}

We have presented the numerical scheme and considered the initial
value problem  in order to stress the dif\/ference of two
equations (\ref{eq:LIE}) and (\ref{eq:LIE_s}). We have found that
the prof\/iles of the vortex f\/ilaments show quite dif\/ferent
behaviour, but for the local stretch, both of the two equations
give the same results in such a way that the initial stretched
region stays in the same place to freeze the stretch or the
shrink.

\LastPageEnding


\begin{thebibliography}{99}
\footnotesize


\bibitem{Konno_Kakuhata2005}
Konno K., Kakuhata H., A new type of stretched solutions excited
by initially stretched vortex f\/ilaments for the local induction
equation, {\it Theor. Math. Phys.}, 2005, V.144, 1181--1189.

\bibitem{Hama_Ames}
Arms R.J., Hama F.R., Localized-induction concept on a curved
vortex and motion of an elliptic vortex ring, {\it Phys. Fluids},
1965, V.8, 553--559.

\bibitem{Konno_Kakuhata2003_2}
Konno K., Kakuhata H., Stretching of vortex f\/ilament with
corrections, in Nonlinear Physics: Theory and Experiment, II
(2002, Gallipoli),
 River Edge, NJ, World Scientif\/ic Publishing, 2003, 273--279.

\bibitem{Hashimoto}
Hashimoto H., A soliton on a vortex f\/ilament, {\it J. Fluid
Mech.} 1972, V.51, 477--485.


\bibitem{Konno_Kakuhata2005_2}
Konno K., Kakuhata H., A hierarchy for integrable equations of
stretched vortex f\/ilament, {\it J. Phys. Soc. Japan}, 2005,
V.74, 1427--1430.

\bibitem{Konno_Mitsuhashi_Ichikawa}
Konno K., Mitsuhashi T., Ichikawa Y.H., Dynamical processes of the
dressed ion acoustic solitons, {\it J. Phys. Soc. Japan}, 1977,
V.43, 669--674.

\end{thebibliography}
\end{document}